\documentclass[aps,prd,preprint,superscriptaddress,tightenlines,nofootinbib]{revtex4}

\usepackage{graphicx}
\usepackage{dcolumn}
\usepackage{bm}

\begin{document}

\preprint{CLEO CONF 06-14}   

\title{Dalitz Analysis of $D^+ \to \pi^+\pi^-\pi^+$}
\thanks{Submitted to the 33$^{\rm rd}$ International Conference on High Energy
Physics, July 26 - August 2, 2006, Moscow}

\author{G.~Bonvicini}
\author{D.~Cinabro}
\author{M.~Dubrovin}
\author{A.~Lincoln}
\affiliation{Wayne State University, Detroit, Michigan 48202}
\author{D.~M.~Asner}
\author{K.~W.~Edwards}
\affiliation{Carleton University, Ottawa, Ontario, Canada K1S 5B6}
\author{R.~A.~Briere}
\author{I.~Brock~\altaffiliation{Current address: Universit\"at Bonn; Nussallee 12; D-53115 Bonn}}
\author{J.~Chen}
\author{T.~Ferguson}
\author{G.~Tatishvili}
\author{H.~Vogel}
\author{M.~E.~Watkins}
\affiliation{Carnegie Mellon University, Pittsburgh, Pennsylvania 15213}
% ...
\collaboration{CLEO Collaboration} 
\noaffiliation
\date{July 24, 2006}

\begin{abstract} 
Using 281/pb of data recorded by the CLEO-c detector observing
$e^+e^-$ collisions at the $\psi(3770)$, corresponding to 1.8 million
$D \bar D$ pairs, the substructure of the decay $D^+ \to \pi^+\pi^-\pi^+$
is investigated using the Dalitz plot technique.
The results presented in this document are preliminary.
\end{abstract}

\maketitle


	A Dalitz plot analysis~\cite{Dalitz} of $D^+ \to \pi^+\pi^-\pi^+$ has previously been done by
E791~\cite{791} and~FOCUS~\cite{focus}.  The preliminary analysis described here is from
CLEO-c~\cite{CLEOdet}, and represents the first time we have done the same Dalitz plot analysis
as the fixed target experiments.  Previously CLEO has focused on analyses with
$\pi^0$'s in the final state.  The decay is selected with cuts on the beam constrained
mass of three charged tracks consistent with pions and the difference of their energy
from the beam energy.  A sample of 4100 events is selected with a signal to noise
of about two to one.  The E791 and FOCUS samples are of similar size and cleanliness.

	The Dalitz plot is symmetric under the interchange of like-sign pions thus we
do the analysis in the two dimensions of high unlike-sign pion mass squared versus
low unlike-sign pion mass.  There is a large contribution from $K_{\rm S}\pi$ which
because of the long $K_{\rm S}$ lifetime should not interfere with the other contributions
to the plot.  We do not consider this stripe on the Dalitz plot when fitting
for two body resonance contributions.
The efficiency across the Dalitz plot is modeled with simulated events that are
fit to a two-dimensional second order polynomial.  While there is a notable
fall of the efficiency in the corners of the Dalitz plot the changes are smooth,
and well modeled by the polynomial.  Backgrounds are taken from sidebands and 
extra resonance contributions to the background are allowed from mismeasured $K_{\rm S}$, $\rho$,
and $f_2(1270)$ decays.  Many possible resonances can contribute to the decay,
and a total of 13 different resonances are considered.  Parameters describing these
resonances are taken from previous experiments.  Only contributions with an
amplitude significant at more than three standard deviations are said to
be observed, and others are limited.  Contributions that are not significant are
not included in the decay model used for the result.

\begin{figure*}[t]
\includegraphics[width=55mm]{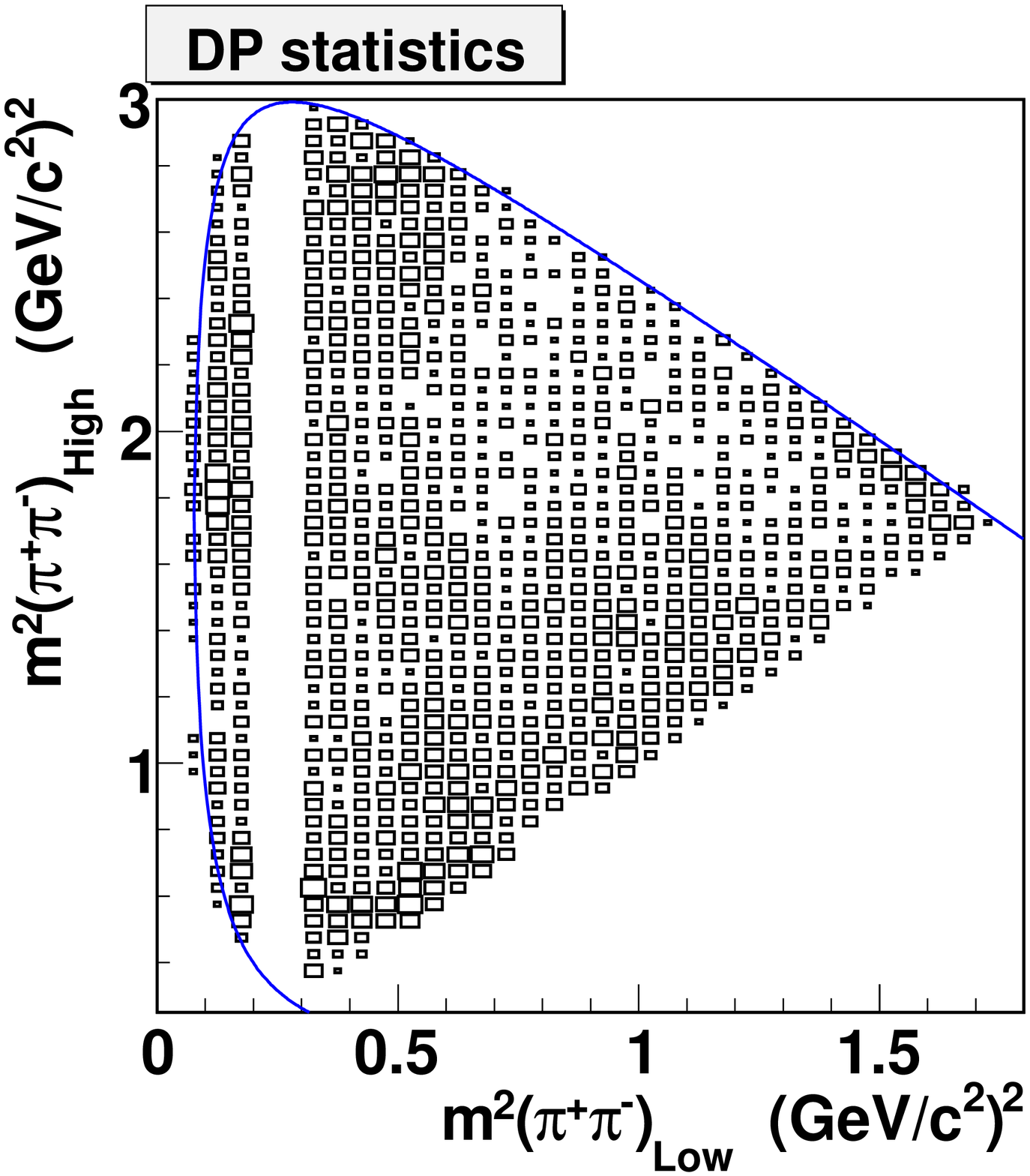}
\includegraphics[width=55mm]{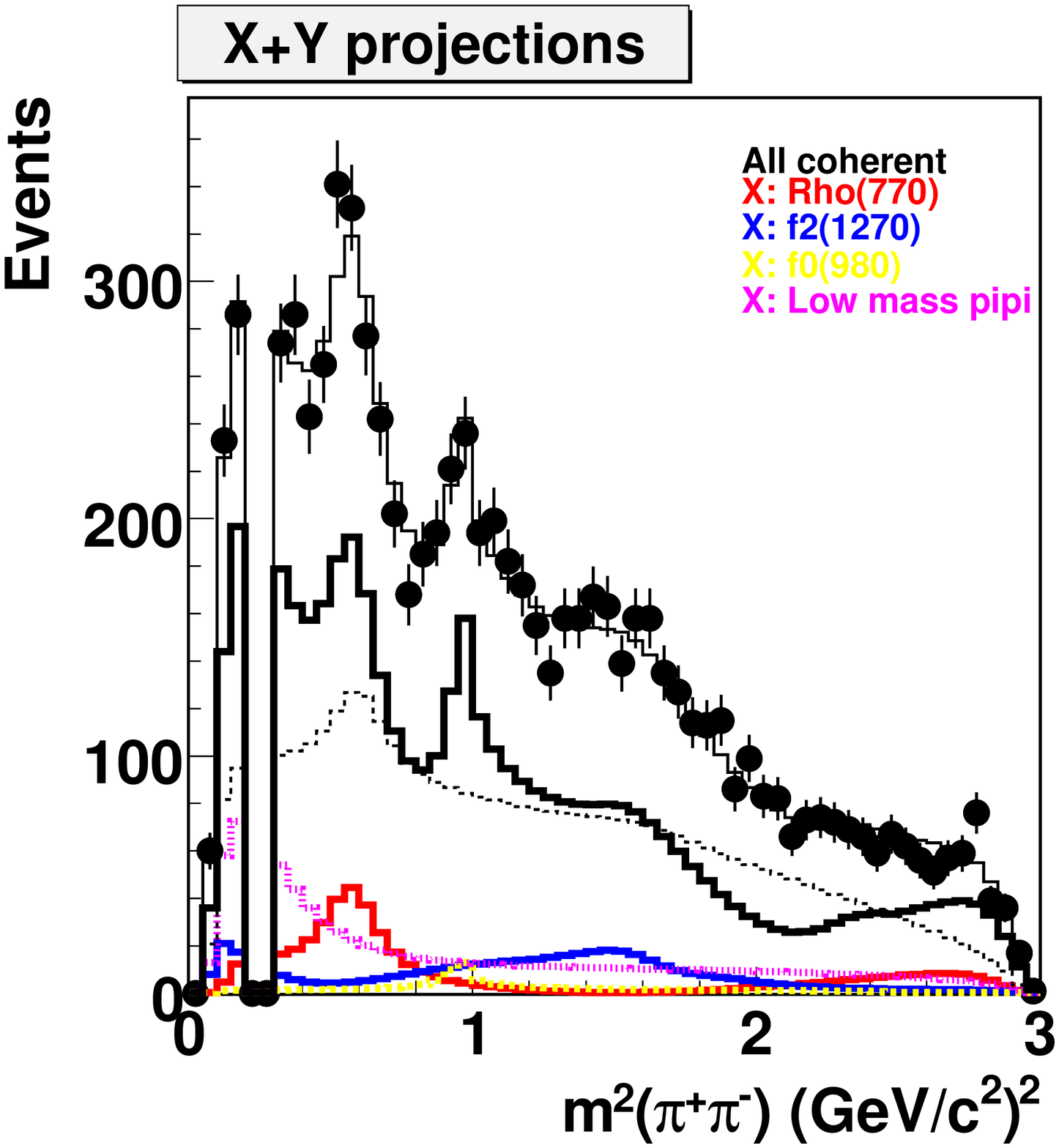}
\includegraphics[width=55mm]{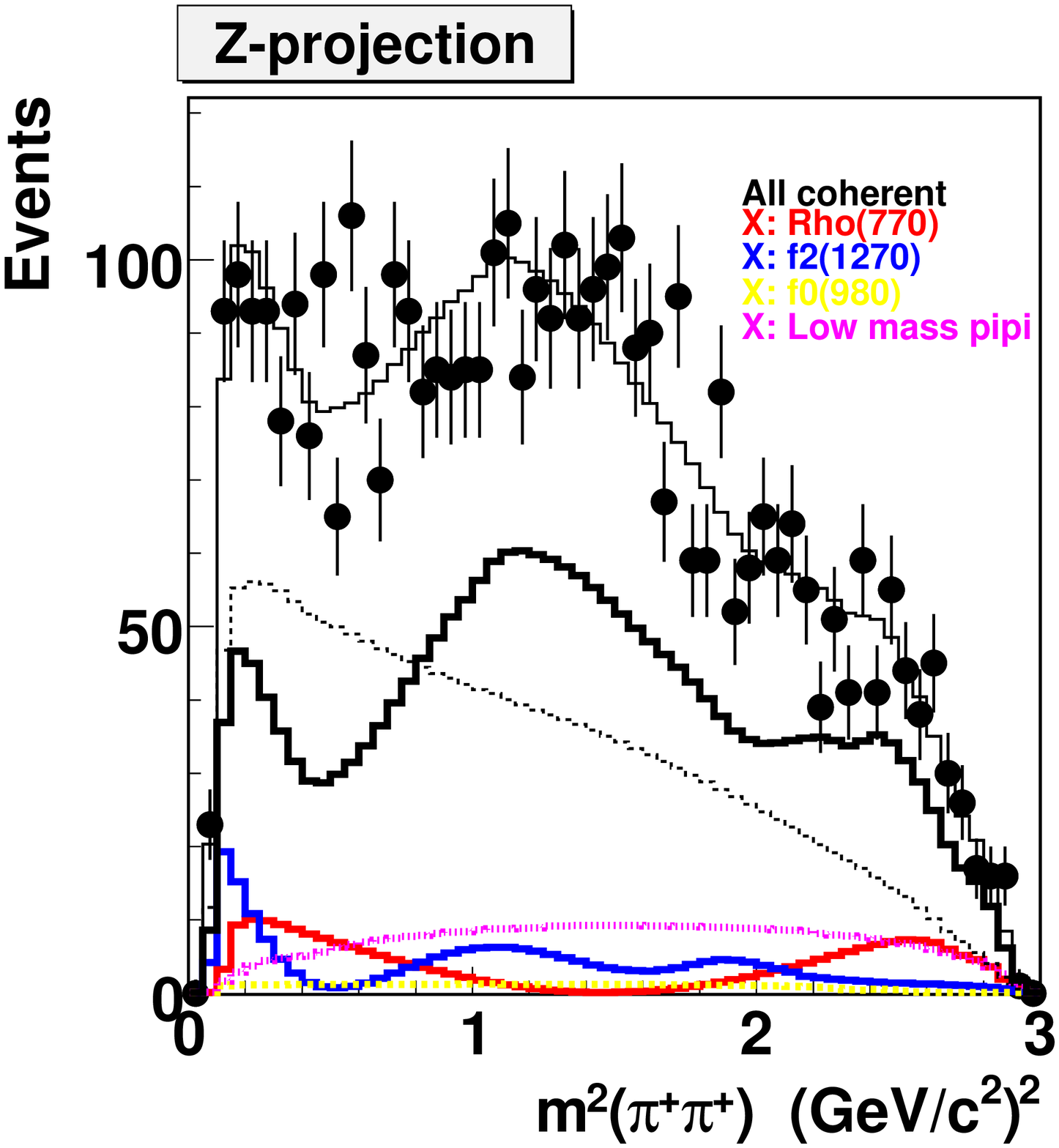}
\caption{Dalitz plot and projections for $D^+ \to \pi^+\pi^-\pi^+$} 
\label{fig:d3pidalitz}
\end{figure*}

	Figure~\ref{fig:d3pidalitz} shows the Dalitz plot and projections on to
the squared masses.  Contributions from $\rho^0\pi$ and $f_2(1270)\pi$ are clearly
visible.  Table~\ref{tab:d3pidalitz} shows the preliminary fit fractions measured by CLEO
\begin{table}[h]
\begin{center}
\caption{Comparison of the fit fractions, in percent, found in the Dalitz plot analysis of $D^+ \to \pi^+\pi^-\pi^+$ between
         the preliminary CLEO analysis and E791.  CLEO limits are at 90\% confidence level.}
\begin{tabular}{|l|c|c|}
\hline 
\textbf{Contribution} & \textbf{CLEO} & \textbf{E791} \\ \hline
$\rho^0 \pi^+$        & $20.0\pm2.5$  & $33.6\pm3.9$ \\
$\sigma^0 \pi^+$      & $41.8\pm2.9$  & $46.3\pm9.2$ \\
$f_2(1270) \pi^+$     & $18.2\pm2.7$  & $19.4\pm2.5$ \\
$f_0(980) \pi^+$      &  $4.1\pm0.9$  & $6.2\pm1.4$ \\
$f_0(1370) \pi^+$     &  $2.6\pm1.9$  & $2.3\pm1.7$ \\
$f_0(1500) \pi^+$     &  $3.4\pm1.3$  & ${}$ \\
Non-resonant          &  $< 3.5$      & $7.8\pm6.6$ \\
$\rho(1450) \pi^+$    &  $< 2.4$      & $0.7\pm0.8$ \\ \hline
\end{tabular}
\label{tab:d3pidalitz}
\end{center}
\end{table}
comparing with the results of the E791 analysis mentioned above.  There is broad
agreement between the two results, including the observation of a $\sigma\pi$
contribution.  In an alternative analysis using the same decay model as E791
the agreement is only slightly better, but the fit is much less likely
to model our data than the model shown in the table which does not include
non-resonant and $\rho(1450) \pi^+$ contributions, but does include a $f_0(1500) \pi^+$
contribution.  Models without a $\sigma\pi$ contribution do not agree well 
with the data.

This CLEO analysis is preliminary, and we plan to consider a generalized
model of $\pi\pi$ S-wave interactions to model $\sigma$ and $f_0$ contributions
such as the K-matrix which is used in the FOCUS analysis mentioned above.

We gratefully acknowledge the effort of the CESR staff
in providing us with excellent luminosity and running conditions.
D.~Cronin-Hennessy and A.~Ryd thank the A.P.~Sloan Foundation.
This work was supported by the National Science Foundation,
the U.S. Department of Energy, and
the Natural Sciences and Engineering Research Council of Canada.

\end{document}